\documentclass[superscriptaddress,dvipsnames,nofootinbib,amsmath,amssymb,prd,notitlepage,twocolumn]{revtex4-1}
\usepackage{color}
\usepackage{bm}
\usepackage{graphicx}
\usepackage{amsmath}
\usepackage{amssymb}
\usepackage{enumitem}
\usepackage{hyperref}
\usepackage{subfigure}
\usepackage{array}
\usepackage[english]{babel}
\usepackage{tensor}
\usepackage{comment}
\usepackage[normalem]{ulem}
\usepackage[dvipsnames]{xcolor}

\def\be{\begin{equation}}
\def\ee{\end{equation}}
\def\ba{\begin{eqnarray}}
\def\ea{\end{eqnarray}}
\def\l{\left}
\def\r{\right}

\allowdisplaybreaks

\begin{document}
\title{$f(Q)$-gravity and neutrino physics}
\author{Lu\'is Atayde}
\email{luisbbatayde@gmail.com}
\affiliation{
 Instituto de Astrofis\'ica e Ci\^{e}ncias do Espa\c{c}o, Faculdade de Ci\^{e}ncias da Universidade de Lisboa, Edificio C8, Campo Grande, P-1749016, Lisboa, Portugal}
\author{Noemi Frusciante}
\email{noemi.frusciante@unina.it}
\affiliation{Dipartimento di Fisica ``E. Pancini", Universit\`a degli Studi  di Napoli  ``Federico II", Compl. Univ. di Monte S. Angelo, Edificio G, Via Cinthia, I-80126, Napoli, Italy}

\begin{abstract}
Within the $f(Q)$-gravity framework we perform a phenomenological study of the  cosmological observables in light of the degeneracy between neutrinos physics and the modified gravity parameter and we identify specific patterns which allow to break such degeneracy. We also provide separately constraints on the total mass of the neutrinos, $\Sigma m_{\nu}$, and on  the effective number of neutrino species, $N_{\rm eff}$, using cosmic microwave background (CMB), baryon acoustic oscillation (BAO), redshift space distortion (RSD), supernovae (SNIa),  galaxy clustering (GC) and weak gravitational lensing (WL)  measurements.  The strongest upper bound  on the total mass of the neutrinos is found for the combination of CMB+BAO+RSD+SNIa and it is $\Sigma m_\nu<0.277$ eV at 95\% C.L. For the same combination of data we find $N_{\rm eff}=2.93^{+0.31}_{-0.34}$ at 95\% C.L. We also find that all combinations of data we consider, prefer a stronger gravitational interaction than $\Lambda$CDM. Finally, we consider the $\chi^2$  and deviance information criterion statistics and find the $f(Q)+\Sigma m_\nu$ model to be  statistically supported by data over the standard scenario. On the contrary $f(Q)+N_{\rm eff}$ is supported by CMB+BAO+RSD+SNIa but a moderate evidence against it is found with GC and WL data.
\end{abstract}

\date{\today}

\maketitle

\section{Introduction}

Extension to the symmetric teleparallel gravity \cite{Nester:1998mp,Adak:2008gd,Adak:2018vzk}  recently got attention from the cosmology community as a possibility to explore new physics beyond the standard cosmological model ($\Lambda$CDM). In this alternative theory, the gravitational interaction is  attributed to the non-metricity scalar $Q$. While an action constructed only with this scalar is equivalent to the one of General Relativity (GR) in flat space \cite{BeltranJimenez:2019esp},  the action  built with a general function of $Q$, i.e. $f(Q)$ \cite{BeltranJimenez:2017tkd,Harko:2018gxr,Xu:2019sbp,BeltranJimenez:2019tme,Jarv:2018bgs,Runkla:2018xrv, BeltranJimenez:2019esp} can generate a gravitational interaction at cosmological scales which shows new interesting patterns both at the level of the background evolution and in the propagation of  scalar and tensor perturbations \cite{BeltranJimenez:2019tme,BeltranJimenez:2019tjy,Lazkoz:2019sjl,Ayuso:2020dcu,Khyllep:2021pcu,Frusciante:2021sio}. Furthermore, for specific forms of the $f(Q)$ function it has  been shown that it can alleviate the $\sigma_8$ tension \cite{Barros:2020bgg}, while others allow for a better fit to cosmological data \cite{Anagnostopoulos:2021ydo,Atayde:2021ujc}.  This theory is then challenging $\Lambda$CDM from different perspectives.

When exploring modified gravity (MG) theories,  such as $f(Q)$-gravity, it is important to consider that MG effects on cosmological observables are highly degenerate  with neutrino physics~\cite{Barreira:2014ija,Shim:2014uta,Baldi:2013iza,He:2013qha,Dossett:2014oia,Hojjati:2011ix,Motohashi:2012wc,Hu:2014sea,Bellomo:2016xhl,Frusciante:2020gkx,Ballardini:2020iws}. 
Indeed, to name a few, a modified gravitational interaction can, for instance, impact the background expansion history, shape   the temperature-temperature cosmic microwave background radiation (CMB) power spectrum at all multipoles ($\ell$) through a modified early and late  integrated Sachs-Wolfe (ISW) effect \cite{Sachs:1967er,Kofman:1985fp} or a change of the amplitude and position of the high-$\ell$ peaks. These effects at high-$\ell$ are due to a modified background expansion history \cite{Hu:1996vq} or to a coupling with dark matter \cite{Amendola:2011ie}. MG can also impact the lensing potential  \cite{Acquaviva:2005xz} and change the growth of structures \cite{Peebles:1984ge,1993MNRAS.262..717B} with a direct effect on the lensing and matter power spectra. Similarly, neutrinos leave detectable signatures on cosmological observations that can be used to constrain their properties, such as the effective number of neutrino species, $N_{\rm eff}$, and the total neutrino mass, $\Sigma m_\nu$ \cite{Lattanzi:2017ubx,TopicalConvenersKNAbazajianJECarlstromATLee:2013bxd}.  Increasing  $N_{\rm eff}$   results in a faster expansion at earlier times altering the radiation to matter equality  which in turn enhances  the first acoustic peak  due to the early ISW effect, while the other CMB peaks are moved to higher multipoles. Additionally it also changes the scale of Silk damping which has the effect of lowering the damping tail of the CMB
spectrum and finally it suppresses  the matter power spectrum. Increasing the sum of the neutrino masses can decrease the low-$\ell$ tail of the  CMB TT power spectrum due to the late ISW effect, change the position of the high-$\ell$ CMB peaks, reduce the weak lensing effect and dump the growth of structures on small scales. Therefore,  constraints on neutrinos properties are highly  dependent on the kind of gravitational interaction one considers. Within the standard cosmological model the upper bound on the sum of the neutrino masses at 95\% for the combination of   Planck15+lensing+BAO+JLA+$H_0$ is $\Sigma m_\nu< 0.23$ eV  \cite{Planck:2015fie} and the determination on the effective number of neutrinos is
 $N_{\rm eff} = 3.15 \pm  0.23$ at  95\% using Planck15+BAO \cite{Planck:2015fie}, the latter being compatible with the prediction of the standard neutrino decoupling model, $N_{\rm eff} =3.046$. Bounds on these two parameters have been found in the MG context as well, for example, when $N_{\rm eff}$ is  fixed to its standard value, one has  for  Ho\v rava gravity  $\Sigma m_\nu<0.165$ eV (95\%, Planck18+lensing+DES+JLA+BKP15) \cite{Frusciante:2020gkx}, for  designer $f(R)$ model $\Sigma m_\nu< 0.32$ eV (95\%, Planck13 +WMAP+BAO+lensing+WiggleZ) \cite{Hu:2014sea},  for the same data set when considering a simple effective field theory of MG model, one has  $\Sigma m_\nu <0.26$ eV \cite{Hu:2014sea}; for the generalized cubic covariant galileon  model instead one is able to find a lower bound $\Sigma m_\nu > 0.11$ eV at 1$\sigma$ using Planck15+BAO+RSD+JLA \cite{Frusciante:2019puu};   for the simplest model of scalar-tensor theories with a conformally coupled scalar field it is $\Sigma m_\nu < 0 .13$ (95\%, Planck18+BAO+HST), while if we vary $N_{\rm eff}$ one finds $3.16\pm 0.19$ at 68\%  and when varying them simultaneously one has $\Sigma m_\nu  < 0 .14$ (95\%) and $N_{\rm eff}=3.14\pm0.20$ (68\%) for the same combinations of data \cite{Ballardini:2020iws}. Finally we mention again the designer $f(R)$ when varying both of them  simultaneously, one finds $N_{\rm eff} = 3.58^{+0.72}_{-
0.69}$ and $\Sigma m_\nu < 0. 860$ eV  (95\%, Planck13+WMAP+BAO+ACT+SPT) \cite{He:2013qha}. One can also notice that when providing bounds on neutrino physics in the presence of modifications of gravity the constraints might be weaker.

In this paper we consider a specific form of the $f(Q)$ function such that the background evolution matches exactly the one of $\Lambda$CDM~\cite{Jimenez:2019ovq}. This model is characterized by one extra free parameter and modification to the gravitational interaction can be appreciated only at perturbation level~\cite{Frusciante:2021sio}.   Cosmological constraints on the modified gravity parameter and on the cosmological parameters have been already derived considering massive neutrinos to have a fixed mass ($\Sigma m_\nu=0.06$ eV, the minimum value allowed
by oscillation experiments) and $N_{\rm eff}=3.046$ ~\cite{Atayde:2021pgb}. In this work we open  the neutrino sector by considering  the mass of the neutrinos or their effective number as additional parameters. We will then be able to investigate whether a degeneracy exists with the modified gravity parameter and if current cosmological data can break such degeneracy.   We will also provide for the first time constraints on $N_{\rm eff}$ and $\Sigma m_\nu$  in the context of $f(Q)$-gravity.

This paper is organized as follows. In Section \ref{Sec:model} we review the theory behind the $f(Q)$-gravity and we present the specific model we analyze. In particular we describe how linear perturbation theory is treated in this analysis. In Sec. \ref{Sec:method} we describe the methodology, the codes and the data sets  we adopt to study the phenomenology associated with the degeneracy between a modified gravitational interaction and the neutrino sector  and to derive the cosmological constraints for model and cosmological parameters, which are thoroughly presented and discussed respectively in Secs.  \ref{Sec:pheno} and \ref{Sec:constraints}. We then provide a model selection analysis in Sec. \ref{Sec:modsel} and finally we conclude in Sec. \ref{Sec:conclusion}.

\section{The model}\label{Sec:model}

 We adopt the Palatini formalism which is characterized  by  the metric, $g_{\mu\nu}$ and the connection $\Gamma^\alpha_{\phantom{X}\mu\nu}$ to be independent fields. The action of the $f(Q)$ theory has the form~\cite{BeltranJimenez:2019tme}:
\be \label{eq:action}
S=\int d^4x\sqrt{-g}\l\{-\frac{1}{16 \pi G_N}\l[Q+f(Q)\r]+\mathcal{L}_m(g_{\mu\nu},\chi_i)\r\}\,,
\ee
where  $g$  is  the determinant of the metric $g_{\mu\nu}$,
$G_N$ is the Newtonian constant and
$Q$ is the non-metricity scalar which is defined as 
 $Q=-Q_{\alpha \mu \nu}P^{\alpha \mu \nu}$ where $Q_{\alpha \mu \nu}$ is the non-metricity tensor, which reads $Q_{\alpha \mu \nu}=\nabla_\alpha g_{\mu \nu}$ and $P^{\alpha}_{\phantom{\alpha}\mu\nu}=-L^{\alpha}_{\phantom{\alpha}\mu\nu}/2+\l(Q^\alpha-\tilde{Q}^\alpha\r)g_{\mu\nu}/4-\delta^\alpha_{(\mu}Q_{\nu)}/4$is the non-metricity coniugate, where we further define $Q_\alpha=g^{\mu\nu}Q_{\alpha\mu\nu}$, $\tilde{Q}_\alpha=g^{\mu\nu}Q_{\mu\alpha\nu}$ and $L^\alpha_{\phantom{\alpha}\mu\nu}=(Q^\alpha_{\phantom{\alpha}\mu\nu}-Q_{(\mu\nu)}^{\phantom{(\mu\nu)} \alpha})/2$. The deviation from GR is encoded in the general function of the non-metricity scalar, i.e. $f(Q)$.  Indeed the action \eqref{eq:action} is equivalent to GR for $f(Q)=0$~\cite{BeltranJimenez:2019tjy}. Finally $\mathcal{L}_m$ is the matter Lagrangian of standard matter fields, $\chi_i$.  Let us note that our action (\ref{eq:action}) when compared with the one in Ref.~\cite{BeltranJimenez:2019tme} includes a redefinition, i.e. $f(Q)\rightarrow \frac{1}{16 \pi G_N}\left(Q+f(Q)\right)$. They are completely equivalent but  this form better fits our purpose. 
 
  We also note that in the following treatment  we  adopt a coordinate choice so that the connection vanishes, following~\cite{BeltranJimenez:2019tme}. This choice is called coincident gauge.

In this work we will consider a flat Friedmann-Lema{\^i}tre-Robertson-Walker (FLRW) metric defined by 
\begin{equation}
    ds^2=-dt^2+a(t)^2\delta_{i}^jdx^idx_j\,,
\end{equation} 
where $t$ is the cosmic time, $x_i$ are the spatial coordinates and $a(t)$ is the scale factor. 

On this background the Friedman equations read~\cite{BeltranJimenez:2017tkd}:
\begin{eqnarray}
  &&    \label{eq:FriedEq}
    H^2(1+ 2 f_Q) - \frac{1}{6} f = \frac{8 \pi G_N}{3} \rho ,  \\
  && (12H^2f_{QQ}+f_Q+1)\dot{H}=-4 \pi G_N(\rho+p)\,,\label{eq:FriedEq2}
\end{eqnarray}
where $H\equiv\dot{a}/a$ is the Hubble parameter and a dot stands for a derivative with respect to cosmic time. We note that on a FLRW background   $Q = 6H^2$ \cite{BeltranJimenez:2017tkd,Jimenez:2019ovq}.   In Eqs. \eqref{eq:FriedEq} and \eqref{eq:FriedEq2} we have also defined:
$f_{Q}\equiv\partial f/\partial Q$, $f_{QQ}\equiv\partial^2 f/\partial Q^2$. Finally
 $\rho$ and $p$  are respectively the sum of the energy density and pressure of non-relativistic matter (baryons and cold dark matter, ''m''), photons (''$\gamma$'') and neutrinos 
  (both massless and massive species, ''$\nu$'').
 Each of them  obeys the continuity equation for perfect fluids, i.e. $\dot{\rho}_i+3H(\rho_i+p_i)=0$.  Baryons and cold dark matter are pressurless fluids, the energy density and the pressure for photons and massless neutrinos are $p_{\gamma,\nu}=1/3\rho_{\gamma,\nu}$. For massive neutrinos we assume a normal hierarchy   obeying the Fermi-Dirac distribution function. 

From the set of background equations, it is clear that either the functional form of $f(Q)$ or the Hubble expansion $H(a)$  needs to be chosen \textit{a priori}. The latter is known as the \textit{designer} approach, firstly applied to $f(R)$ theory \cite{Song:0610532,Pogosian:0709} and later to $f(Q)$ theory as well \cite{Albuquerque:2022eac}. In this investigation we want to consider a background evolution that matches the one of the standard cosmological model, $\Lambda$CDM, in this way it is possible to solve analytically the Friedmann equation for $f(Q)$.  Then one gets~\cite{Jimenez:2019ovq}:
\be\label{model}
f(Q) =  \alpha H_0\sqrt{Q} + 6H_0^2 \Omega_\Lambda,
\ee
where the parameter $\alpha$ is a dimensionless constant, $H_0\equiv H(a=1)$ is the Hubble parameter evaluated at present time ($a=1$) and  $\Omega_\Lambda$ is the energy density parameter of the cosmological constant. 
This construction then allows to investigate deviations from $\Lambda$CDM that appear at linear perturbation level. 

A perturbed flat FLRW metric can be written in Newtonian gauge as follows
 \be
 ds^2=-(1+2\Psi)dt^2+a^2(1-2\Phi)\delta_{i}^jdx^idx_j\,,
 \ee 
 where  $\Phi(t,x_i)$ and $\Psi(t,x_i)$ are the two gravitational potentials. In GR these two potentials are equal but for modified gravity theories they can be different. Then the relations between the two gravitational potentials and the gauge-invariant density contrast, $\Delta$, are altered compared to GR.  
 We adopt the following general way to parameterize modified gravity in Fourier space, which consists of introducing two coupling functions, $\mu(a,k)$ and $\gamma(a,k)$, which are defined through the following equations~\cite{Amendola:2007rr,PhysRevD.81.083534,Silvestri:2013ne,2010PhRvD..81j4023P,Amendola:2019laa}:
\begin{eqnarray}\label{eq:Poisson}
-\frac{k^2}{a^2}\Psi&=& 4 \pi G_N\mu (a,k)[\rho\Delta+3(\rho+p)\sigma] ,\\
\frac{k^2}{a^2}[\phi -\gamma(a,k)\Psi] &=& 12\pi G_N\mu(a, k)(\rho + p) \sigma\,.
\end{eqnarray}
Here we define 
 the gauge-invariant density contrast as
 \begin{equation}
   \rho\Delta\equiv \rho\delta+3\frac{aH}{k^2}(\rho+p)  \,,
 \end{equation}
  with   $\delta$ being the density contrast,  $\sigma$ is the matter anisotropic stress and $k$ is the Fourier mode.  In particular $\mu$ is the
 effective gravitational coupling and accounts for the modifications of gravity on the clustering of matter. Having $\mu>1$ implies a stronger gravitational interaction than in GR ($\mu_{GR}=1$), while a weaker gravity is obtained for $\mu<1$. Instead $\gamma$ is known as the slip parameter but it does not have a direct connection to observables. 
 For this reason usually one considers another coupling function $\Sigma(a,k)$, known as the light deflection parameter, which describes how light travels on cosmological
distances. It is defined via the following equation~\cite{Amendola:2007rr,PhysRevD.81.083534,Silvestri:2013ne,2010PhRvD..81j4023P,Amendola:2019laa}:
\begin{eqnarray}
\frac{k^2}{a^2}(\Phi+\Psi)=-4\pi G_N\Sigma(a,k)[2\rho\Delta+3(\rho+p)\sigma].  
\end{eqnarray}
In the limiting case of negligible matter anisotropic stress the three coupling functions are connected: 
\begin{equation}
  \Sigma=\frac{\mu}{2}(1+\gamma)\,.  
\end{equation}

 In this approach  the coupling functions  $\mu$, $\Sigma$ and $\gamma$ in general can depend on both scale and time and $\gamma\neq1$, i.e. $\phi\neq\psi$. However, when computing analytically their forms in specific theories of MG, by using the quasi static approximation (QSA) \cite{Boisseau:2000pr,DeFelice:2011hq},  they show up to be scale independent, e.g.  Brans Dicke theory \cite{Brans:1961sx}, DGP \cite{Lombriser:2013aj}, K-mouflage \cite{Benevento:2018xcu}, Galileon/Horndeski theories  \cite{DeFelice:2011hq}. On the contrary $f(R)$ models  show an explicitly scale dependence \cite{Pogosian:2007sw}. Additionally  some of them have $\gamma=1$, e.g. Cubic Galileon models and DGP, which seems a feature connected with an unchanged speed of propagation of tensor modes with respect to GR \cite{Pogosian:2016pwr}.  Let us note that the validity of the QSA for  dark energy and modified gravity models has been extensively discussed in Ref. \cite{ Sawicki:2015zya} and numerically proved to be valid for some specific models for modes with $k \gtrsim 0.001 $ h/Mpc in Refs. \cite{Pogosian:2016pwr,Frusciante:2019xia}.

In the case of $f(Q)$-gravity, it is possible to obtain  the equations to linear order ~\cite{Jimenez:2019ovq}, which 
  assuming the QSA, leads to  $\gamma=1$, similarly  to GR and to a modified Poisson equation \eqref{eq:Poisson} with~\cite{Jimenez:2019ovq}: 
 \be
\mu(a)=\frac{1}{1+f_Q}\,,
\ee
which is scale independent. We note that due to stability requirement $1+f_Q>0$.
For the model we consider in this work (Eq. \eqref{model}) it reads~\cite{Frusciante:2021sio}:
\be\label{eq:mu}
\mu(a)= \frac{12 H}{12H+\sqrt{6}\alpha H_0} \,.
\ee
Such modification in the strength of the gravitational interaction has been shown to lead to significant and measurable signatures on cosmological observables compared to $\Lambda$CDM \cite{Frusciante:2021sio}: when $\alpha>0$ the gravitational interaction is weaker than in GR and this leads to a suppressed  lensing potential auto-correlation
power spectrum and matter power spectrum and to an enhanced CMB ISW-tail; the opposite holds in the case of  $\alpha<0$ which corresponds to a stronger gravitational interaction.  Let us note that the model we are analysing does not have early time modifications  since the $f(Q)$ function reduces to its $\Lambda$CDM limit at early times, showing deviations from standard model only at $z+1\lesssim 20$ \cite{Frusciante:2021sio}.

\section{Methodology and data sets}\label{Sec:method}

In this work we are interested in studying the degeneracy between a modified gravitational interaction given by the $f(Q)$ gravity and the neutrino sector.   For this purpose we adopt the Einstein-Boltzmann code \texttt{MGCAMB} \cite{Hojjati:2011ix,Zucca:2019xhg} which implements the framework for linear perturbations described in Section \ref{Sec:model} and has an accurate modelling of massive neutrinos \cite{Zucca:2019xhg}. In detail we use a modified version of it \cite{Frusciante:2021sio} which  includes a new patch to evolve the model in Eq. \eqref{model}.  For specific details about the explicit modifications of the equations in \texttt{MGCAMB} on the inclusion of massive neutrinos in the context of MG and the $f(Q)$ model under consideration we refer the reader to \cite{Zucca:2019xhg} and \cite{Frusciante:2021sio} respectively. We will then use the Monte-Carlo Markov Chain code, \texttt{MGCosmoMC} \cite{Hojjati:2011ix} to provide cosmological constraints. The sampled chains from \texttt{MGCosmoMC}
are analyzed using \texttt{GetDist} \cite{Lewis:2019xzd}.

 We will present in the following Sections both the phenomenology analysis and the cosmological constraints which will be splitted into two cases:
\begin{enumerate}
    \item investigation of the impact of varying massive neutrinos. For this case we will assume three 
   neutrino mass eigenstates  $m_1$, $m_2$, and $m_3$ with $m_3 \gg m_2 > m_1$ (normal hierarchy) 
     and a fixed $N_{\rm eff}=3.046$.
    \item  investigation of the impact of varying the neutrino effective number, $N_{\rm eff}$, due to  neutrinos and to any additional massless particles.
\end{enumerate}

Let us now present the sets of data we will use for the cosmological constraints.

We use  the Planck 2018 \cite{Planck:2019nip} data of CMB temperature likelihood for large angular scales ($\ell =[2,29]$ for TT power spectrum) and for the small angular scales a joint  TT, TE and EE likelihoods ($\ell =[30,2508]$ for TT power spectrum, $\ell =[30,1996]$ for TE cross-correlation and EE power spectra). We refer to this data set as ''Plk18''. 

We also consider baryonic acoustic oscillation (BAO) data from the Sloan Digital Sky Survey (SDSS) DR7 Main Galaxy Sample \cite{Ross:2014qpa} and 
the 6dF Galaxy Survey \cite{Beutler:2011hx} and 
 the joint BAO and RSD datasets from the SDSS DR12 consensus release \cite{BOSS:2016wmc}.

 Additionally we include the Joint Light-curve Array (JLA) of Supernova Type IA (SNIa) from the  Supernova Legacy Survey (SNLS) and SDSS \cite{SDSS:2014iwm}.
We refer to the combined analysis of Planck18, BAO, RSD and SNIa data as ''PBRS''. 

The latest data set we consider includes galaxy clustering (GC) and weak gravitational lensing (WL) measurements from the Dark Energy Survey Year-One (DES-1Y) data \cite{DES:2017myr}.
 In details these are measurements of the angular two-point correlation functions of GC, cosmic shear and galaxy-galaxy lensing. We note that for the present analysis we adopt a standard DES cutoff  (by setting $\Delta \chi^2 = 5$) of the non-linear regime  following \cite{Planck:2015bue,DES:2018ufa,Zucca:2019xhg} because for $f(Q)$-gravity a model of non-linearities  is not yet available. We name the joint analysis of DES data with the previous data sets as ''PBRSD''.

Additional details about the analysis include  a flat prior on the model's parameter $\alpha \in [-3,3]$ and we vary the base cosmological parameters such as the physical densities of baryons  ($\Omega_b h^2 \in [0.005,0.1]$, where $h=H_0/100$) and cold dark matter ($\Omega_c h^2 \in [0.001, 0.99]$),  the primordial amplitude ($\ln(10^{10} A_s) \in [1.61, 3.91]$), the reionization optical depth ($\tau \in [0.01, 0.8]$), the angular size of the sound horizon at recombination ($\theta_{MC} \in [0.5, 10]$) and spectral index  of scalar perturbations ($n_s \in [0.8, 1.2]$).  Depending on which of the two cases we are considering we also vary either $\Sigma m_\nu \in [0.01, 0.9]$ or $N_{\rm eff}\in [1.0, 5.0]$.   We will show the results in terms of $\sigma_8^0$ (which is a derived parameter) in place of $A_s$.

\section{Phenomenology: degeneracy with the neutrino sector}\label{Sec:pheno} 

 In this Section we will investigate the degeneracy between the MG parameter, $\alpha$, and, alternately, with $\Sigma m_\nu$ and $N_{\rm eff}$. 
To this purpose we will show the effects of these parameters on some cosmological observables such as the CMB TT, lensing and matter power spectra. We will use the following values for the cosmological  parameters:
$\Omega_c  h^2=0.12011$,
$\Omega_b  h^2=0.02238$,
$H_0=67.32 $ km/s/Mpc,
$A_s= 3.0448 \times 10^{-9}$ and
$n_s=0.966$. For the MG parameter we consider two cases: $\alpha=0.5$ and $\alpha=-0.5$. Finally
when studying the impact of the total mass of  neutrinos we have $\Omega_\nu h^2=0.008602$ with $\Sigma m_\nu=0.8$ eV and $\Omega_\nu h^2=0.000645$ with $\Sigma m_\nu=0.06$ eV and we fix $N_{\rm eff}=3.046$. When investigating the degeneracy with $N_{\rm eff}$ we select $N_{\rm eff}=3.046$ and  $N_{\rm eff}=3.8$ and massive neutrinos is set to zero. Let us stress that these values are purely illustrative.

\begin{figure}[ht!]
    \centering
\includegraphics[width=0.44\textwidth]{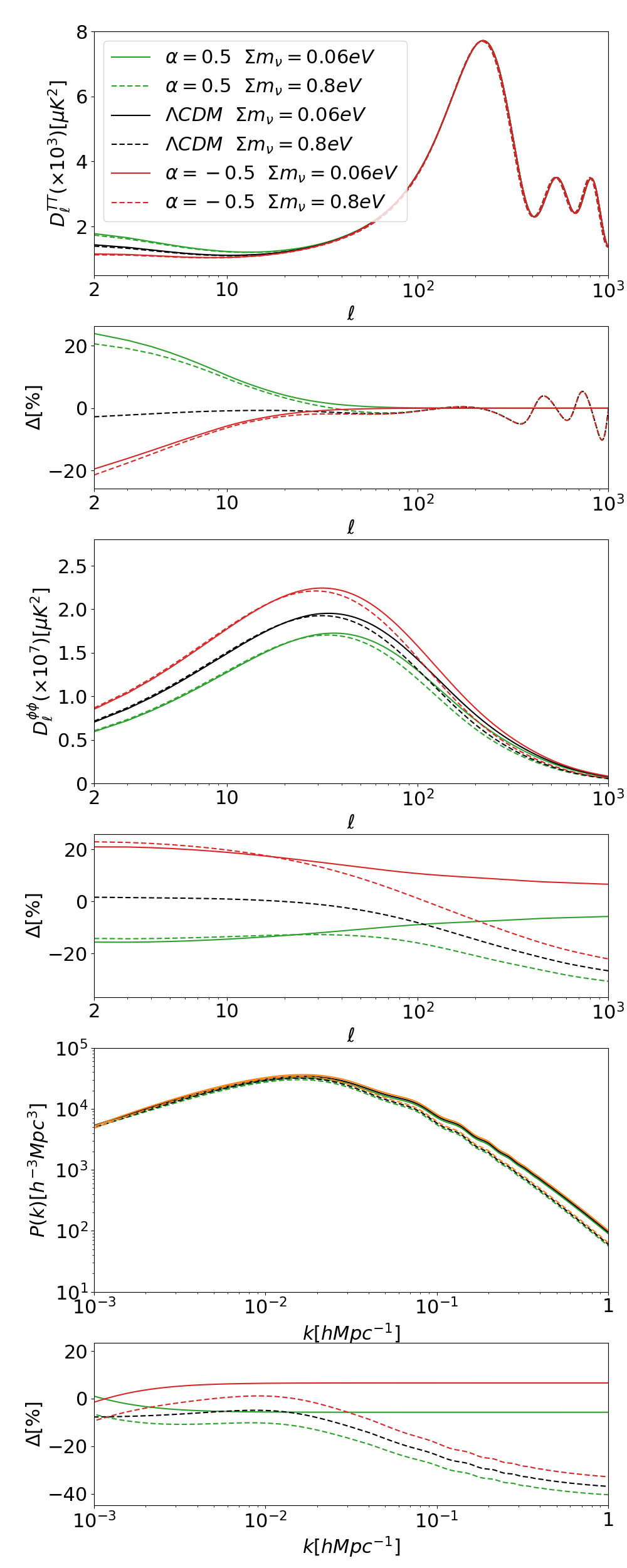}
    \caption{CMB TT power spectrum ($D^{TT}_\ell =\ell (\ell + 1)C^{TT}_\ell/(2\pi)$, upper panel),   lensing potential auto-correlation power spectra ($D^{\phi\phi}_\ell = \ell(\ell + 1)C^{\phi\phi}_\ell/(2\pi)$, middle panel) and matter power spectra $P(k)$ (bottom panel) for $f(Q)$-gravity and $\Lambda$CDM when two values for $\Sigma m_\nu$ are chosen. For each of them we provide the percentage relative difference ($\Delta$) with respect to the $\Lambda$CDM model with $\Sigma m_\nu=0.06$ eV.}
\label{fig:Phenomenology_fig_vn}
\end{figure}

\begin{itemize}

\item \textit{The neutrino mass}

In Figure \ref{fig:Phenomenology_fig_vn} we show the impact on the cosmological observables of varying the total mass of neutrinos and the modified gravity parameter. These have been already discussed in the Introduction and Sec. \ref{Sec:model} respectively. Let us now focus on their degeneracy. 

The specific modification of gravity we are considering impacts the CMB TT power spectrum at low-$\ell$ (top panels) as an increase  of the sum of neutrino mass does. In particular we can notice that a weaker gravitational interaction ($\alpha>0$) has the opposite effect of a larger $\Sigma m_\nu$: while the former enhances the CMB ISW tail, the latter suppresses it, implying that the values of these two parameters can be fine-tuned to  cancel their effects. Thus showing the existence of a degeneracy between them.
When the gravitational interaction is stronger ($\alpha<0$) the ISW-tail is suppressed by 
modified gravity and the inclusion of massive neutrinos will further suppress the power spectrum at low-$\ell$. This means that though the two effects cannot cancel each other, they can be fine-tuned to obtain a given CMB TT power spectrum at low-$\ell$. What can make a difference to break the degeneracy in the CMB TT power spectrum between these two parameter is the impact $\alpha$ and $\Sigma m_\nu$ have on the CMB TT power spectrum at high-$\ell$. The former does not change its shape because the background is the same of $\Lambda$CDM, while the higher is the total mass of the neutrinos the larger is the  shift of the acoustic peaks to higher multipoles. This is an important aspect because data sets which are able to constrain the background evolution can be used to strongly constrain  $\Sigma m_\nu$, thus mitigating its degeneracy with $\alpha$.

In the two middle panels of Figure \ref{fig:Phenomenology_fig_vn}, we show the effects of varying  $\Sigma m_\nu$ and $\alpha$ on the lensing potential auto-correlation
power spectrum and we also show the relative percentage difference  with
respect to the $\Lambda$CDM model with $\Sigma m_\nu = 0.06$ eV. An increase of the neutrino mass reduces the lensing effect as it is well known \cite{Lattanzi:2017ubx,TopicalConvenersKNAbazajianJECarlstromATLee:2013bxd}, similarly to the effect a positive $\alpha$ has.  On the contrary a negative  value of $\alpha$ increases it. Let us note that differently of what happens in the ISW tail of the CMB spectrum, here the  direction of the degeneracies goes in the opposite direction, i.e. while  in the ISW tail of CMB spectrum an effect due to a positive $\alpha$ can be compensated by increasing $\Sigma m_\nu$, in the lensing power spectrum they will both contribute in suppressing its power. This is another clear pattern that can contribute in breaking the degeneracy between these two parameters.

Finally we discuss the degeneracy that appears in the matter power spectrum (bottom panels). Increasing the total mass of neutrinos damps the matter power spectrum on large $k$, same effect generated by a weak gravitational interaction. Instead a stronger gravitational interaction enhances the matter power spectrum. We note that a compensation of the MG effect and that of an increasing $\Sigma m_\nu$ might be hard to obtain due to the scale dependence introduced by the massive neutrinos.  A scale dependence that is not present in our  $f(Q)$-gravity.  Let us note that while here we focus on massive neutrinos and $\alpha$ degeneracy, other degeneracies between $\alpha$/$\Sigma m_\nu$ and other cosmological parameters might exist.

In summary we have identified and discussed degeneracy effects between $\alpha$ and $\Sigma m_\nu$ but we have also highlighted those patterns that can allow us to break their degeneracies.
\vspace{1cm}

\begin{figure}[ht!]
    \centering
\includegraphics[width=0.44\textwidth]{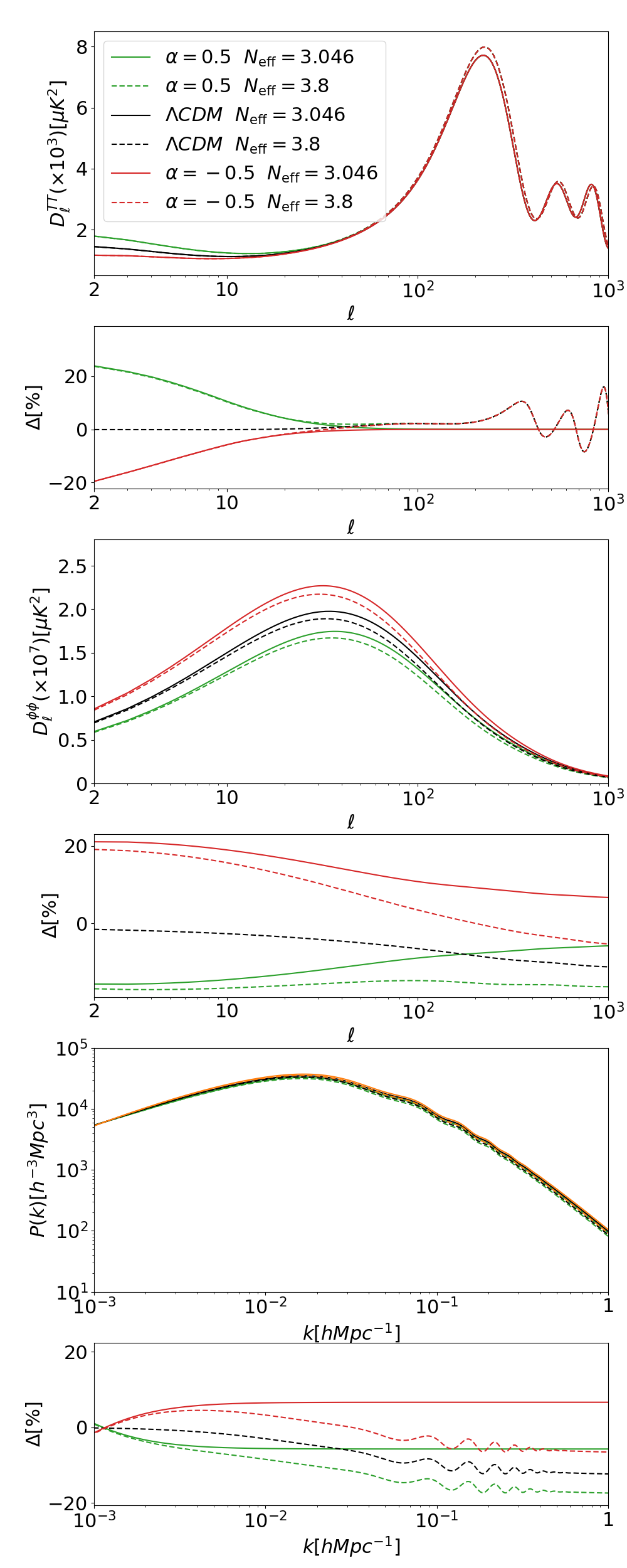}
    \caption{CMB TT power spectrum ($D^{TT}_\ell =\ell (\ell + 1)C^{TT}_\ell/(2\pi)$, upper panel),   lensing potential auto-correlation power spectra ($D^{\phi\phi}_\ell = \ell(\ell + 1)C^{\phi\phi}_\ell/(2\pi)$, middle panel) and matter power spectra $P(k)$ (bottom panel) for $f(Q)$-gravity and $\Lambda$CDM when two values for $N_{\rm eff}$ are chosen. For each of them we provide the percentage relative difference ($\Delta$) with respect to the $\Lambda$CDM model with $N_{\rm eff}=3.046$.}
    \label{fig:Phenomenology_fig_nnu}
\end{figure}

\begin{figure*}[th!]
    \centering
\includegraphics[width=0.95\textwidth]{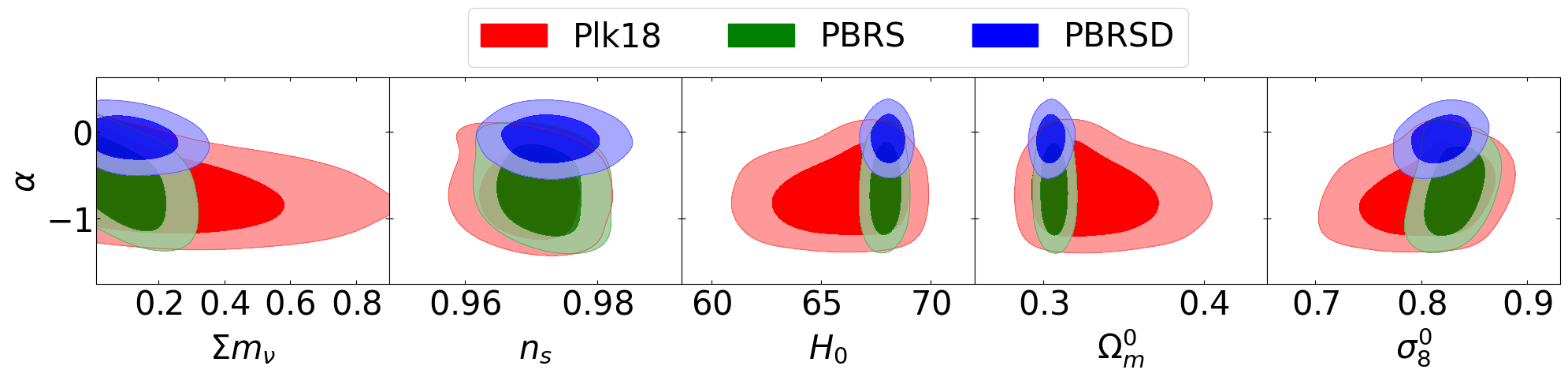}
     \caption{Marginalised constraints at 68\% (darker) and 95\% (lighter) C.L. on the model parameter $\alpha$ and five cosmological parameters $\Sigma m_{\nu}$, $n_s$, $H_0$, $\Omega_{m}^0$ and $\sigma_8^0$  obtained with the CMB data from Planck 2018 (PLK18, red), its combination with BAO, RSD and SNIa data (PBRS, green) and with DES data (PBRSD, blue).}
    \label{fig:FQ_mnu_All}
\end{figure*}
\item \textit{The neutrino effective number}

In Figure \ref{fig:Phenomenology_fig_nnu} we show the effects of varying $N_{\rm eff}$ along with $\alpha$ on some cosmological observables.

In the top two panels we can notice that modified gravity and an increasing $N_{\rm eff}$ have completely different effects on the CMB spectrum: while the former affects the amplitude of the ISW-tail at low-$\ell$ due to a late time ISW effect, the latter impacts the shape and positions of the acoustic peaks at high-$\ell$ due to a combined effect of the early ISW effect and background cosmology. 

In the two middle panels we show the effects on the lensing power spectrum. A higher value of $N_{\rm eff}$ suppresses the lensing spectrum and, as we already saw before, this is a similar effect produced by a weaker gravitational interaction, contrary to what happen when the gravitational interaction is stronger. While here some degeneracy might exist we need to consider that the dependence of  $N_{\rm eff}$ and MG from the angular scale is different as it is evident in the plot of the percentage relative difference. 

Finally in the bottom two panels we show the matter power spectrum and the percentage relative difference. Here we find the same situation discussed for the case of massive neutrinos: a degeneracy exists but the scale dependence and the damping of the acoustic scale introduced by a larger value of $N_{\rm eff}$ make the difference. 

In summary the degeneracy between MG and $N_{\rm eff}$ appears to be less relevant than the one identified for the case of $\Sigma m_\nu$.

\end{itemize}

\begin{figure*}[t!]
    \centering
    \includegraphics[width=0.95\textwidth]{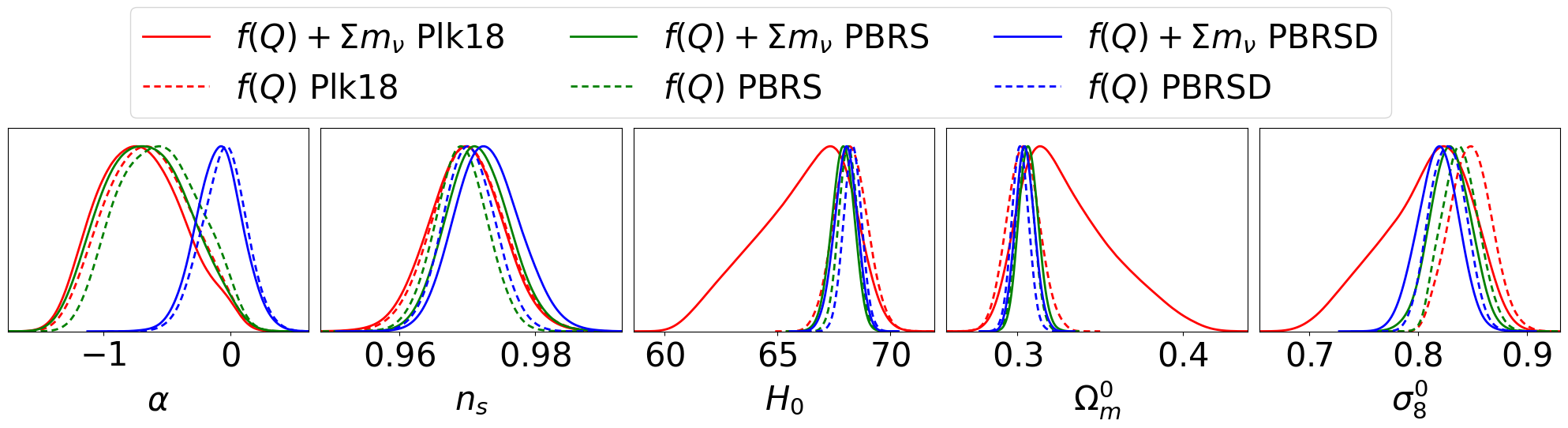}  \includegraphics[width=0.95\textwidth]{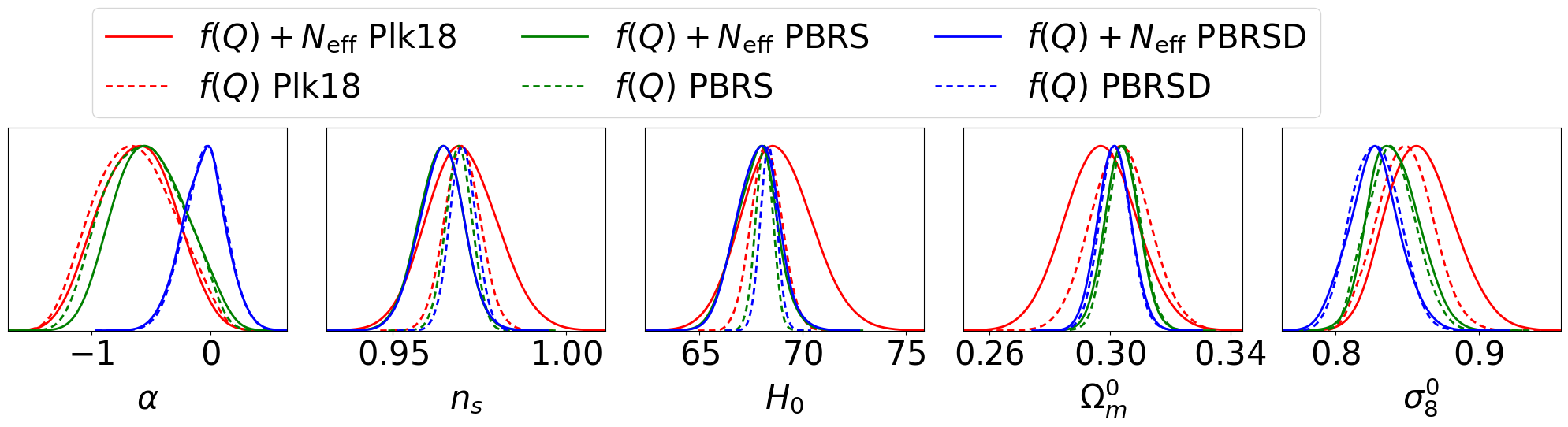}
    \caption{Top panel: marginalized cosmological parameters in the case of $f(Q)$-gravity with varying massive neutrinos (solid lines, results in this paper) and the case with fixed total mass of neutrinos $\Sigma m_\nu=0.06$ eV in Ref. \cite{Atayde:2021pgb} (dashed lines). Bottom panel: marginalized cosmological parameters in the case of $f(Q)$-gravity with varying $N_{\rm eff}$ (solid lines, results in this paper) and the case of fixed $N_{\rm eff}=3.046$ in Ref. \cite{Atayde:2021pgb} (dashed lines).}
    \label{fig:FQ_mnu_1D}
\end{figure*}

\section{Cosmological constraints} \label{Sec:constraints}

In this Section we present the cosmological constraints on the model parameter and cosmological parameters for the cases of varying the total mass of neutrinos and the neutrino effective number. They can be found respectively in Tables \ref{Tab:VaryMassiveNeu} and \ref{Tab:VaryNeff} which also include the  constraints for the respective cases in the $\Lambda$CDM scenario for comparison.

\begin{figure*}[t!]
    \centering  
\includegraphics[width=0.9\textwidth]{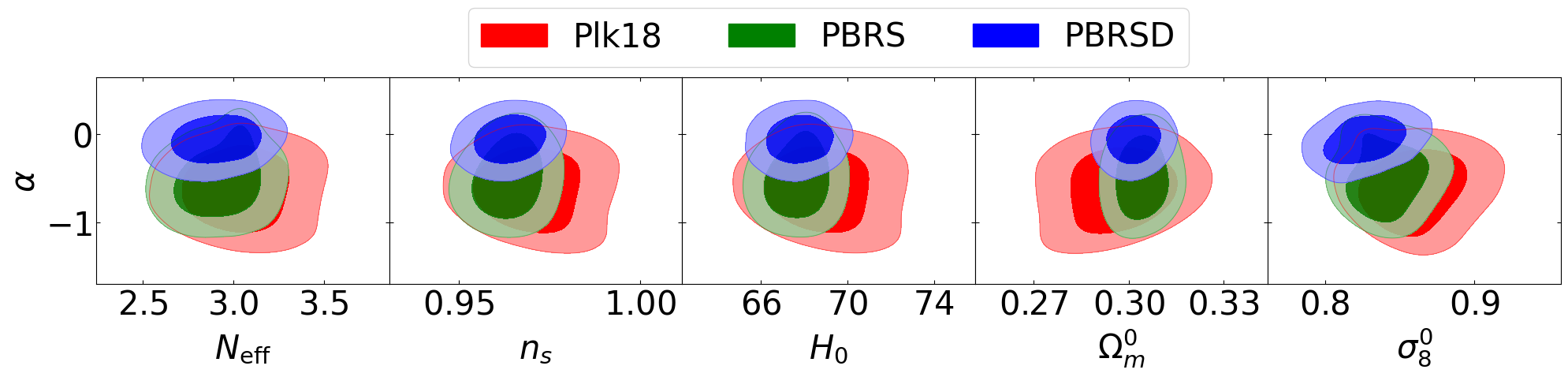}
    \caption{Marginalised constraints at 68\% (darker) and 95\% (lighter) C.L. on the model parameter $\alpha$ and five cosmological parameters $N_{\rm eff}$, $n_s$, $H_0$, $\Omega_{m}^0$ and $\sigma_8^0$  obtained with the CMB data from Planck 2018 (PLK18, red), its combination with BAO, RSD and SNIa data (PBRS, green) and with DES data (PBRSD, blue).}
    \label{fig:FQ_n_eff_All}
\end{figure*}

\begin{itemize}
\item \textit{The neutrino mass}

We now discuss the cosmological constraints for the $f(Q)+\Sigma m_\nu$ case. From Fig. \ref{fig:FQ_mnu_All} we can notice that the stronger upper bound on $\Sigma m_\nu$ is found for the PBRS data ($\Sigma m_\nu<0.277$ eV at 95\% C.L.) which is driven by the power in constraining  background data have on $\Omega_m^0$. Regardless of the combination of data we note that for MG the upper bounds are weaker than the corresponding cases in  $\Lambda$CDM. In particular we note that for the full combination of data the bound degrades a bit because DES data prefer a weaker gravitational interaction, while CMB data for the $f(Q)$ model support a stronger gravitational interaction. Though the value of $\alpha$ is higher for this data combination with respect to the one for Plk18, it is not sufficient to obtain a weaker gravitational interaction which is then supplied by a large total mass of neutrinos.  Looking at the constraints on the cosmological parameters we note they are compatible within the error with those of the standard scenario. The constraints on the $\alpha$ parameter are in line with what is found in Ref. \cite{Atayde:2021ujc}. In detail a larger negative mean value of $\alpha$ is preferred by the Plk18 because it allows for a better fit of the low-$\ell$ tail of TT power spectrum. On the contrary a higher value is preferred by DES data. In Fig. \ref{fig:FQ_mnu_1D} top panel we compare our results with those obtained for the same model but fixed $\Sigma m_\nu=0.06$ eV \cite{Atayde:2021ujc}. We note that they are compatible within the errors but the $f(Q)+\Sigma m_\nu$ case in general prefers smaller mean values for $\alpha$, this is because the effects of $\alpha$ are then compensated by a larger $\Sigma m_\nu$. We also notice that the mean values of $\sigma_8^0$  and $H_0$ are slightly lower for the varying massive neutrinos case.

\begin{table*}
\centering
\begin{tabular}{|l|l|l|l|l|l|l|l|}
\hline
\multicolumn{8}{|c|}{Varying massive neutrinos }  \\ \hline\hline
 Model &  $\alpha $ & ${\rm{ln}}(10^{10} A_s)$ & $n_s$ & $H_0$ & $\Omega_m^0  $ & $\sigma_8^0    $ & $\Sigma m_\nu $\\ \hline \hline
 $\Lambda$CDM  (Plk18)& - & $3.17^{+0.10}_{-0.12}      $ & $0.97\pm0.01   $ &    $66.6^{+2.9}_{-4.1}        $ & $0.325^{+0.054}_{-0.038}   $ & $0.823^{+0.064}_{-0.080}   $ & $< 0.622                   $ \\
 $\Lambda$CDM  (PBRS)& -  & $3.14^{+0.09}_{-0.10}    $ & $0.970\pm 0.008$& $67.97^{+0.87}_{-0.91}     $ & $0.306\pm 0.011   $ & $0.839\pm 0.035  $ & $< 0.219                   $ \\
 $\Lambda$CDM  (PBRSD)& - &   $3.13^{+0.10}_{-0.10}      $& $0.973^{+0.010}_{-0.0093}  $& $68.08^{+0.93}_{-1.0}      $& $0.304^{+0.012}_{-0.011}   $ & $0.822^{+0.035}_{-0.034}   $& $< 0.277                   $ \\ \hline
 $f(Q)$ (Plk18)& $-0.70^{+0.65}_{-0.60}$ & $3.09 \pm 0.13  $ & $0.970\pm0.010  $ & $66.0^{+3.4}_{-4.2}        $ &  $0.332^{+0.057}_{-0.044}   $ & $0.806^{+0.070}_{-0.084}   $ & $< 0.731  $          \\
 $f(Q)$ (PBRS)& $-0.66^{+0.66}_{-0.60}     $ & $3.06^{+0.12}_{-0.11}      $ & $0.972\pm 0.009$ & $67.92^{+0.93}_{-1.0}      $ & $0.307^{+0.012}_{-0.011}   $ & $0.830^{+0.036}_{-0.037}   $  & $< 0.277                   $ \\
 $f(Q)$ (PBRSD)& $-0.09^{+0.36}_{-0.36}     $ & $3.12^{+0.11}_{-0.11}      $ & $0.9731^{+0.0098}_{-0.0089}$ & $68.04^{+0.95}_{-1.0}      $&  $0.305^{+0.012}_{-0.011}   $& $0.819^{+0.036}_{-0.037}   $& $< 0.293                   $       \\ \hline
\end{tabular}
\caption{Marginalized constraints on cosmological and model parameters at 95\% C.L. for the $\Lambda$CDM and $f(Q)$ models when the total mass of the neutrinos is considered as free parameter. }
\label{Tab:VaryMassiveNeu}
\end{table*}
\vspace{1cm}

\item \textit{The neutrino effective number}\label{Sec:conneff}

In Fig. \ref{fig:FQ_n_eff_All} we show the cosmological constraints for the $f(Q)+N_{\rm eff}$ case.
The bounds on $N_{\rm eff}$ are always compatible within the errors   with the prediction of the standard neutrino decoupling model. We note that regardless of the combination of data the $f(Q)+N_{\rm eff}$ case has slightly lower mean values for $N_{\rm eff}$ compared to the ones for the $\Lambda$CDM$+N_{\rm eff}$ case.
The constraints on the other  cosmological parameters  are very similar to  those of the standard scenario. When varying $N_{\rm eff}$ we notice that the mean values of $\alpha$ are higher compared to those we obtain for $f(Q)+\Sigma m_\nu$ or those discussed in Ref. \cite{Atayde:2021ujc} for fixed $N_{\rm eff}$ and $\Sigma m_\nu$ (see Fig. \ref{fig:FQ_mnu_1D} bottom panel for comparison)   because higher values of $\alpha$ are necessary to compensate the lower value of $N_{\rm eff}$.

\end{itemize}

\begin{table*}
\centering
\begin{tabular}{|l|l|l|l|l|l|l|l|}
\hline
\multicolumn{8}{|c|}{Varying the neutrino effective number }  \\ \hline\hline
 Model &  $\alpha $ & ${\rm{ln}}(10^{10} A_s)$ & $n_s$ & $H_0$ & $\Omega_m^0  $ & $\sigma_8^0    $ & $N_{eff}        $\\ \hline \hline
 $\Lambda$CDM  (Plk18)& - & $3.14^{+0.11}_{-0.11}      $ & $0.971^{+0.022}_{-0.020}   $ &    $68.7^{+3.6}_{-3.3}        $     & $0.300^{+0.024}_{-0.025}   $ & $0.862^{+0.052}_{-0.048}    $ & $3.07^{+0.43}_{-0.40}      $ \\
 $\Lambda$CDM  (PBRS)& -  & $3.106^{+0.082}_{-0.084}   $ & $0.966^{+0.013}_{-0.013}   $ & $67.9^{+2.0}_{-2.0}        $ & $0.304^{+0.011}_{-0.011}   $ & $0.846^{+0.039}_{-0.037}   $ & $2.98^{+0.32}_{-0.33}      $\\
 $\Lambda$CDM  (PBRSD)& - &   $3.072^{+0.080}_{-0.080}   $ &  $0.964^{+0.012}_{-0.013}   $ & $67.8^{+1.9}_{-2.1}        $ & $0.302^{+0.011}_{-0.011}   $ &  $0.828^{+0.035}_{-0.034}   $ &  $2.91^{+0.29}_{-0.34}      $ \\ \hline
 $f(Q)$ (Plk18)& $-0.62^{+0.59}_{-0.59}     $ & $3.06^{+0.13}_{-0.11}      $ &  $0.970^{+0.021}_{-0.019}   $ & $68.8^{+3.5}_{-3.2}        $ &  $0.297^{+0.023}_{-0.023}   $ & $0.860^{+0.047}_{-0.044}   $ & $3.03^{+0.42}_{-0.39}      $          \\
 $f(Q)$ (PBRS)& $-0.50^{+0.59}_{-0.54}     $& $3.03^{+0.10}_{-0.091}     $ & $0.964^{+0.013}_{-0.013}   $ &  $67.7^{+2.0}_{-2.1}        $ & $0.304^{+0.011}_{-0.011}   $ & $0.841^{+0.036}_{-0.032}   $ & $2.93^{+0.31}_{-0.34}      $ \\
 $f(Q)$ (PBRSD)& $-0.06^{+0.36}_{-0.37}     $ & $3.061^{+0.099}_{-0.099}   $ & $0.964^{+0.012}_{-0.013}   $ & $67.8^{+2.0}_{-2.0}        $&  $0.302^{+0.011}_{-0.011}   $& $0.827^{+0.036}_{-0.034}   $& $2.91^{+0.31}_{-0.33}      $       \\ \hline
\end{tabular}
\caption{Marginalized constraints on cosmological and model parameters at 95\% C.L. for the $\Lambda$CDM and $f(Q)$ model  when $N_{\rm eff}$ is considered as free parameter. }
\label{Tab:VaryNeff}
\end{table*}

\section{Model selection analysis}\label{Sec:modsel}

We provide a model selection analysis based on the $\chi^2$-statistics and  
 the  Deviance Information Criterion (DIC)~\cite{RSSB:RSSB12062}.  While the former allows us to measure how a model compares to actual observed data, the latter 
 will allow us to  quantify the preference of one model over the other and it is particulary useful when the posterior distributions of the models have been obtained by MCMC simulation. In particular in both cases we will compute the difference  of the  $f(Q)$ model with respect to $\Lambda$CDM: $\Delta \chi^2= \chi^2_{\rm f(Q)}-\chi^2_{\Lambda {\rm CDM}}$ and $\Delta \text{DIC} = \text{DIC}_\text{f(Q)} - \text{DIC}_\text{$\Lambda$CDM}$. Before moving to the actual discussion of results, let us introduce the  definition we use to compute the DIC, which is
\be
\text{DIC}:= \chi_\text{eff}^2 + 2 p_\text{D},
\ee
where $\chi_\text{eff}^2$ is the value of the effective $\chi^2$ corresponding to the maximum likelihood and $p_\text{D} = \overline{\chi}_\text{eff}^2 - \chi_\text{eff}^2$, with the bar being the average of the posterior distribution. This definition allows us to take into account  both the goodness of fit  and the bayesian complexity of the model, according to which  more complex models are disfavored. Therefore models with smaller DIC are preferred to models with larger DIC \cite{Liddle:2009xe,Peirone:2019aua,Peirone:2019yjs,Frusciante:2019puu,Frusciante:2020gkx,Anagnostopoulos:2021ydo,Rezaei:2021qpq}. 
The  values for both the $\Delta \chi^2$ and  $\Delta {\rm DIC}$ for the case of varying the mass of neutrinos are shown in Table \ref{tab:mnu} and for the case of varying the effective number are Table \ref{tab:neff}. 

In the case of $f(Q)+\Sigma m_{\nu}$ we find that, regardless of the combination of data, the  model can fit the data better than $\Lambda$CDM$+\Sigma m_{\nu}$ according to the $\chi^2$-statistics. This is mostly driven by a better fit to CMB data. Additionally according to the \text{DIC} the model is also statistically preferred by data specifically for the Plk18 and PBRS data sets, while no evidence is found when GC and WL data are included. However let us note that the $\Delta \text{DIC}$ between $f(Q)+\Sigma m_{\nu}$ and $\Lambda$CDM$+\Sigma m_{\nu}$, when computed for the full combination, gets better if compared to the one found between $f(Q)$ and $\Lambda$CDM for a fixed $\Sigma m_{\nu}$, $\Delta \text{DIC}=4.8203 
$ \cite{Atayde:2021ujc}. In the case under consideration  a large mass of the neutrinos can suppress the matter power spectrum and compensate the enhancement due to the stronger gravitational interaction which is preferred by CMB data.

In the case of $f(Q)+N_{\rm eff}$, the  $\chi^2$-statistics tells us that the model fits better  the data with respect to $\Lambda$CDM$+N_{\rm eff}$ for all combinations.  The $\Delta \text{DIC}$ supports the model over the standard scenario  when considering CMB, BAO, RSD and SNIa, however this is not the case when we include DES data. That is because, though the value of $N_{\rm eff}$ is lowered, it cannot compensate the higher value of $\alpha$ preferred by DES data which degrades the CMB fit.

In summary the model selection analysis we performed shows a support for the  
$f(Q)$ model over $\Lambda$CDM  for the CMB, BAO, RSD and SNIa data.  When DES data are included we do not find a definite conclusion when we consider $f(Q)+\Sigma m_{\nu}$ which needs to be further explored, possibly including non-linear scales, and a moderate evidence against $f(Q)+N_{\rm eff}$ ($\Delta \text{DIC} > 5$ according to Jeffrey’s scale \cite{Kass:1995loi}).

\begin{table}[t!]
\centering
\begin{tabular}{|l|l|l|l|l|l|l|l|}
\hline
 \multicolumn{3}{|c|}{Varying massive neutrinos }  \\ \hline\hline
 Data & $\Delta \chi^2$ &$\Delta DIC$ 
  \\ \hline
  Plk18 & -2.284 & -6.547 \\
PBRS & -2.938 & -3.987 \\
PBRSD & -2.470  & 0.732 \\ \hline
\end{tabular}
\caption{ Results for the $\Delta \chi^2=\chi^2_{\rm f(Q)}-\chi^2_{\Lambda {\rm CDM}}$ and $\Delta {\rm DIC}={\rm DIC}_{\rm f(Q)}-{\rm DIC}_{\Lambda {\rm CDM}}$. }\label{tab:mnu}
\end{table}

\begin{table}[t!]
\centering
\begin{tabular}{|l|l|l|l|l|l|l|l|}
\hline
 \multicolumn{3}{|c|}{Varying $N_{\rm eff}$}  \\ \hline\hline
 Data & $\Delta \chi^2$ &$\Delta DIC$ 
  \\ \hline
Plk18 & -5.282 & -1.975 \\
PBRS & -3.204 & -3.153 \\
PBRSD & -2.468 & 4.985 \\ \hline
\end{tabular}
\caption{ Results for the $\Delta \chi^2=\chi^2_{\rm f(Q)}-\chi^2_{\Lambda {\rm CDM}}$ and $\Delta {\rm DIC}={\rm DIC}_{\rm f(Q)}-{\rm DIC}_{\Lambda {\rm CDM}}$. }
\label{tab:neff}
\end{table}

\section{Conclusions}\label{Sec:conclusion}

We studied the degeneracy in the cosmological observables between either the total mass of neutrinos or the effective number of neutrino species  with the modification of gravity introduced by an $f(Q)$ model which is characterized by a background evolution identical to the $\Lambda$CDM one. In this way we limited the degeneracy at the linear perturbation level only.  We found that  degeneracy exists and that the one between MG and $\Sigma m_\nu$ is more relevant than the one present with $N_{\rm eff}$. In particular we noticed that such degeneracies can be mitigated by a combination with background data which will strongly constrain either $N_{\rm eff}$ and $\Sigma m_\nu$ since the MG effects enter only at linear perturbation level given the choice we made for the form of the $f(Q)$ function. 

We also put observational constraints on the model parameter $\alpha$ and the other cosmological parameters, separately considering the case of varying $\Sigma m_\nu$ and $N_{\rm eff}$ by adopting MCMC simulation with CMB, BAO, RSD, SNIa, GC and WL measurements. The strongest upper bound on $\Sigma m_\nu$ is obtained with the combination of CMB, BAO, RSD, SNIa with the constraining power coming from BAO, RSD, SNIa. $N_{\rm eff}$ is always found to be compatible  with the standard value of $N_{\rm eff}=3.046$ but with lower mean values.  
The  best-fit values of the $f(Q)$ model gave for all combinations of data a better effective $\chi^2$-statistic than the standard cosmological scenario. When we considered the DIC statistic we noted that the combinations of Plk18 alone and the one with BAO, RSD, SNIa were those for which we noted a statistical support by data. When the combination includes DES data we did not find an evidence in support or against $f(Q)$ in the case of varying $\Sigma m_\nu$ but we found a moderate evidence against $f(Q)+N_{\rm eff}$. Thus the $f(Q)+\Sigma m_\nu$ model  can be a compelling and viable candidate for explaining the late time acceleration. 

In the future we will perform further investigations which will consider a background for $f(Q)$ which is not exactly $\Lambda$CDM. Of interest will be also the inclusion of non-linear scales in modeling the GC and WL so that we can fully consider DES data.

\acknowledgments
We thank Bruno J. Barros and Francesco Pace for useful comments on the results.
L.A. is supported by Fundação para a Ciência e a Tecnologia (FCT) through the research grants UIDB/04434/2020, UIDP/04434/2020 and  from the FCT PhD fellowship grant with ref. number 2022.11152.BD .
N.F. is supported by the Italian Ministry of University and Research (MUR) through the Rita Levi Montalcini project ``Tests of gravity on cosmic scales" with reference PGR19ILFGP.
The authors also acknowledge the FCT project with ref. number PTDC/FIS-AST/0054/2021 and  the COST Action CosmoVerse, CA21136, supported by COST (European Cooperation in Science and Technology).

\bibliography{sample}

\end{document}